\documentstyle[sprocl,epsfig]{article}

\def\gfvh{g_{fV\h}}
\def\gfvphi{g_{fV\phi}}

\def\Hhat{\widehat H}
\def\vo{v_0}
\def\ho{h_0}
\def\mho{m_{\ho}}
\def\phio{\phi_0}
\def\mphio{m_{\phio}}

\def\NIM{\em Nucl. Instrum. Methods}

\def\NPB{{\em Nucl. Phys.} B}
\def\PLB{{\em Phys. Lett.}  B}
\def\PRL{\em Phys. Rev. Lett.}
\def\PRD{{\em Phys. Rev.} D}

\def\be{\begin{equation}}
\def\ee{\end{equation}}
\def\bea{\begin{eqnarray}}
\def\eea{\end{eqnarray}}

\def\gfvh{g_{fV\h}}
\def\gfvphi{g_{fV\phi}}

\def\Hhat{\widehat H}
\def\vo{v_0}
\def\ho{h_0}
\def\mho{m_{\ho}}
\def\phio{\phi_0}
\def\mphio{m_{\phio}}
\def\h{h}
\def\mh{m_{h}}
\def\mphi{m_\phi}

\def\lphi{\Lambda_\phi}

\def\mphi{m_\phi}

\def\hbar{\overline h}

\def\mpl{M_{Planck}}

\def\h{h}
\def\mh{m_{\h}}

\def\lsim{\mathrel{\raise.3ex\hbox{$<$\kern-.75em\lower1ex\hbox{$\sim$}}}}
\def\gsim{\mathrel{\raise.3ex\hbox{$>$\kern-.75em\lower1ex\hbox{$\sim$}}}}
\def\ifmath#1{\relax\ifmmode #1\else $#1$\fi}
\def\half{\ifmath{{\textstyle{1 \over 2}}}}

\def\mplanck{M_{\rm Planck}}
\def\mpl{\mplanck}

\def\eg{{\it e.g.}}

\def\eg{{\it e.g.}}

\def\hsm{h_{\rm SM}}
\def\mhsm{m_{\hsm}}
\def\hsm{h_{SM}}
\def\mhsm{m_{\hsm}}

\def\mw{m_W}

\def\MPL #1 #2 #3 {{\sl Mod.~Phys.~Lett.}~{\bf#1} (#3) #2}
\def\NPB #1 #2 #3 {{\sl Nucl.~Phys.}~{\bf #1} (#3) #2}
\def\PLB #1 #2 #3 {{\sl Phys.~Lett.}~{\bf #1} (#3) #2}
\def\PR #1 #2 #3 {{\sl Phys.~Rep.}~{\bf#1} (#3) #2}
\def\PRD #1 #2 #3 {{\sl Phys.~Rev.}~{\bf #1} (#3) #2}
\def\PRL #1 #2 #3 {{\sl Phys.~Rev.~Lett.}~{\bf#1} (#3) #2}
\def\RMP #1 #2 #3 {{\sl Rev.~Mod.~Phys.}~{\bf#1} (#3) #2}
\def\ZPC #1 #2 #3 {{\sl Z.~Phys.}~{\bf #1} (#3) #2}
\def\IJMP #1 #2 #3 {{\sl Int.~J.~Mod.~Phys.}~{\bf#1} (#3) #2}
\def\NIM #1 #2 #3 {{\sl Nucl.~Inst.~and~Meth.}~{\bf#1} {#3} #2}

\def\gam{\gamma}

\def\anti{\overline}
\def\epem{e^+e^-}

\def\eg{{\it e.g.}}

\def\anti{\overline}

\def\fbi{~{\rm fb}^{-1}}

\def\gev{~{\rm GeV}}
\def\tev{~{\rm TeV}}

\newcommand{\nc}{\newcommand}
\nc{\beq}{\begin{equation}}   \nc{\eeq}{\end{equation}}
\nc{\baa}{\begin{array}}      \nc{\eaa}{\end{array}}
\nc{\bit}{\begin{itemize}}    \nc{\eit}{\end{itemize}}
\nc{\ben}{\begin{enumerate}}  \nc{\een}{\end{enumerate}}
\nc{\bce}{\begin{center}}     \nc{\ece}{\end{center}}
\def\beqa{\begin{eqnarray}}
\def\eeqa{\end{eqnarray}}
\def\bed{\begin{description}}
\def\eed{\end{description}}

\def\eg{{\it e.g.}}

\def\ee{e^+e^-}

\def\gamc{$\gam$C}

\begin{document}

\title{The Need for a Photon-Photon Collider 
in addition to LHC \& ILC for Unraveling
the Scalar Sector of the Randall-Sundrum Model~\footnote{To appear
  in the Proceedings of the International Conference on Linear Colliders, Paris, April 19-23, 2004.}
\vspace*{-.17in}}

\author{JOHN F. GUNION}

\address{Department of Physics, University of California at
  Davis, Davis CA 95616 }


\maketitle\abstracts{\vspace*{-.1in} 
  In the Randall-Sundrum model there can be a rich new phenomenology
  associated with Higgs-radion mixing.  A photon-photon collider
  (\gamc) would provide a crucial complement to the LHC and future ILC
  colliders for fully determining the parameters of the model and
  definitively testing it.  
\vspace*{-.17in}}
  
First, I review the essential features of the Randall-Sundrum (RS)
model \cite{rs}.
There are two branes, separated in the 5th dimension, $y$, and
$y\to -y$ symmetry is imposed.  With appropriate boundary conditions,
the 5D Einstein equations yield the metric
\beq
ds^2=e^{-2\sigma(y)}\eta_{\mu\nu}dx^\mu dx^\nu-b_0^2dy^2, 
\label{metricz}
\eeq
where $\sigma(y)\sim m_0b_0|y|$.
Here, $e^{-2\sigma(y)}$ is the warp factor which reduces
scales of order $\mpl$ at $y=0$ 
on the hidden brane to scales of order a TeV at $y=1/2$ on the
visible brane.
Fluctuations of $g_{\mu\nu}$ relative to $\eta_{\mu\nu}$ 
are the KK excitations $h^n_{\mu\nu}$.
Fluctuations of $b(x)$ relative to $b_0$ define the radion field.
In addition, we place a Higgs doublet $\Hhat$ on the visible brane.
After various rescalings, the properly normalized radion and Higgs quantum fluctuation
fields are denoted by $\phio$ and $h_0$.
The action responsible for Higgs-radion mixing \cite{wellsmix} is
\beq
S_\xi=\xi \int d^4 x \sqrt{g_{\rm vis}}R(g_{\rm vis})\Hhat^\dagger \Hhat\,,
\eeq
where $R(g_{\rm vis})$ is the Ricci scalar 
for the metric induced on the visible brane.

\begin{figure}[h!]
\begin{center}
\includegraphics[width=3.5in]{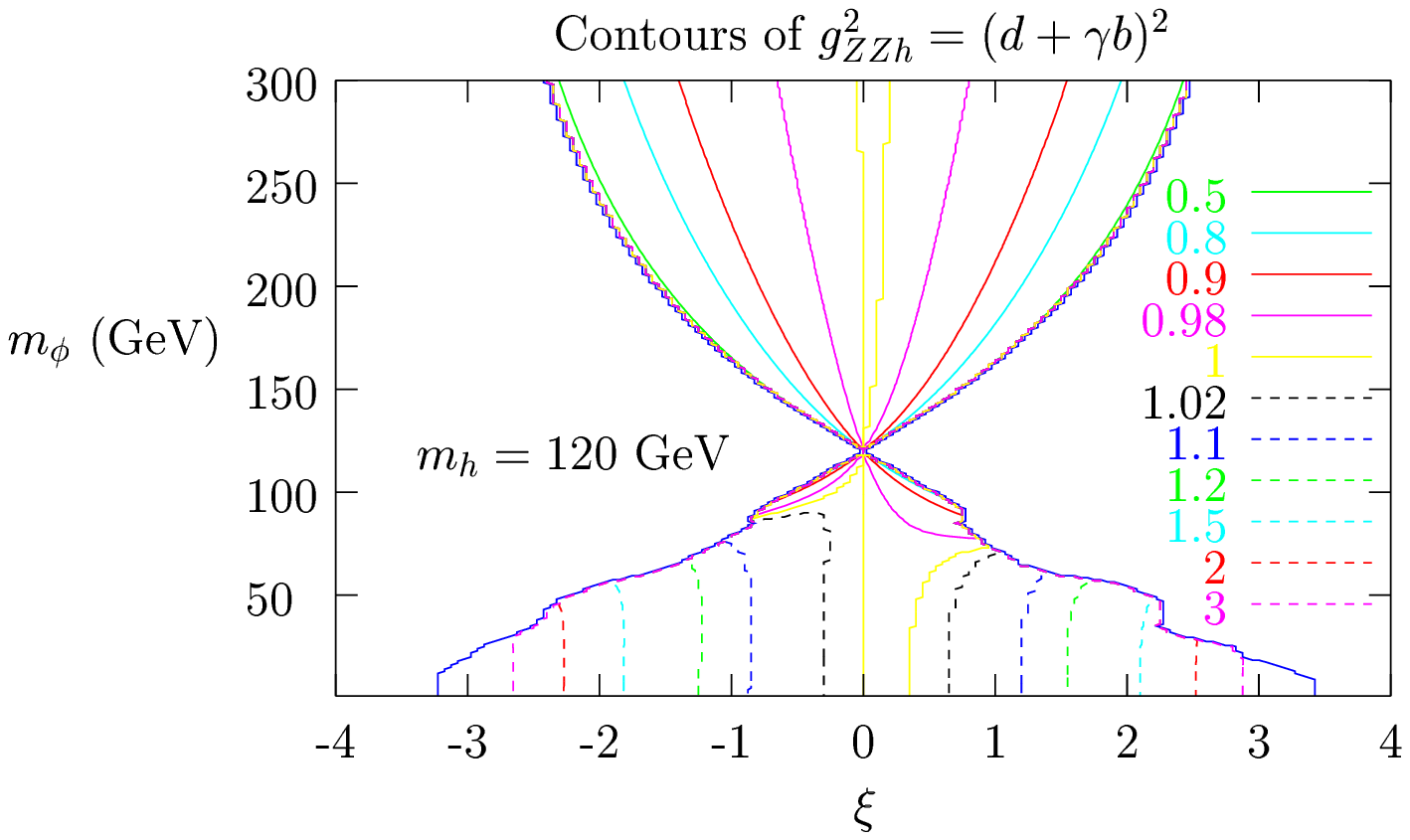}
\vspace*{.1in}
\includegraphics[width=3.5in]{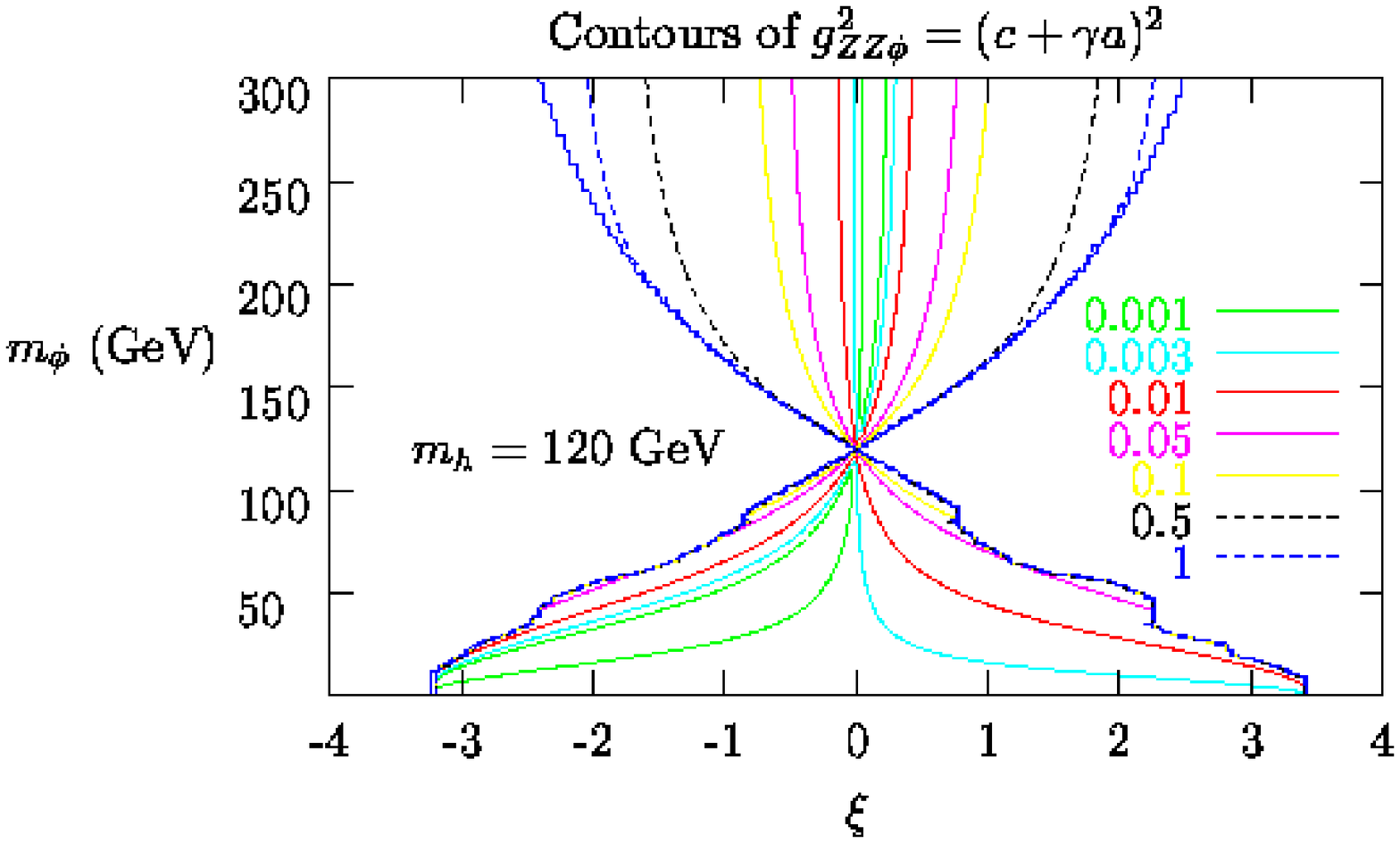}
\end{center}\vspace*{-.15in}
\caption{Contours of $g_{ZZh}^2=\gfvh^2$ and
  $g_{ZZ\phi}^2=\gfvphi^2$  [defined relative to the SM Higgs, Eq.~(\ref{gdefs})]
for $\lphi=5\tev$ and $\mh=120\gev$. Regions outside the
hour-glass shape are theoretically inconsistent. LEP direct discovery
limits have been imposed and (at large $|\xi|$) cut out parts of the otherwise
allowed $\mphi<\mh$ parameter region. From~\protect\cite{Dominici:2002jv}.}
\label{couplingshmh120}
\vspace*{-.15in}
\end{figure}

A crucial parameter is the ratio
$\gamma\equiv \vo/\lphi$ where $\vo=246\gev$ is the SM Higgs vev and
$\lphi$ is the vacuum expectation value of the radion field.
The full quadratic structure of the Lagrangian,
including $\xi\neq 0$ mixing, takes a form in which
the $\ho$ and $\phio$ fields for $\xi=0$ are mixed and have complicated
kinetic energy normalization.
We must diagonalize and rescale to get the canonically
normalized mass eigenstate fields, $h$ and $\phi$ \cite{csakimix}:
\bea
\ho
&\equiv& d\h+c\phi
\qquad \phio
\equiv a\phi+b\h\,. \label{phiform}
\eea
In the above equations, $a,b,c,d$ are functions of $\xi$,
$\gam$ and the bare masses, $\mho$ and $\mphio$. 
For given values of $\xi$, $\gam$,
$\mh$ and $\mphi$, one must invert a set of equations
to determine $\mho$ and $\mphio$ and, thence, $a,b,c,d$.
Requiring consistency leads to strong constraints on
the allowed $\xi$ values for fixed $\mh$, $\mphi$ and $\gam$,
leading to an hourglass shape for the theoretically
allowed region in $(\xi,\mphi)$ parameter space at fixed $\mh$ and
fixed $\gam$ (equivalently, fixed $\lphi$), as shown in Fig.~\ref{couplingshmh120}
\cite{Dominici:2002jv,Hewett:2002nk}. The precision EW
studies of Ref.~\cite{Gunion:2003px} 
suggest that some of the larger $|\xi|$
range is excluded, but we studied the whole range just in case.

The KK-graviton couplings to the $h$ and $\phi$ are determined by $\lphi$.
Fortunately, $\lphi$ can be extracted using measurements
of the KK-graviton spectrum at the LHC.  In particular,
the mass of the first KK-excitation is given by
$m_1=x_1 {m_0\over\mpl} {\lphi\over\sqrt 6}$,
where $x_1$ is the first
zero of the Bessel function $J_1$ ($x_1\sim 3.8$), while the excitation spectrum as a function of
$m_{jj}$ in the vicinity of $m_1$ determines $m_0/\mpl$ (see, for
example, the plots in
\cite{Davoudiasl:1999jd}). The ratio
$m_0/\mpl$ is related to the curvature of the brane
and should be a relatively small number for consistency
of the RS scenario. Sample parameters that are safe from precision EW data
and RunI Tevatron constraints are \cite{Davoudiasl:1999jd}
$\lphi=5\tev$ and $m_0/\mpl=0.1$ (the latter is employed for all plots presented).
These give $m_1\sim 780\gev$,~\footnote{Note that $m_1$ is typically too large
for KK graviton excitations 
to be present, or if present, important, in $h,\phi$ decays.} well
within the LHC reach. Once $\lphi$ is determined, the goal will be to extract 
$\xi$, $\mh$ and $\mphi$  from Higgs-radion measurements.

Crucial to determining these model parameters are
the $f\anti f$ and $VV$ couplings of the $\h$ and $\phi$.
For $V=W,Z$ and all $f$, the 
$h$ and $\phi$ couplings are rescaled relative to SM $\hsm$ couplings by 
the universal factors $\gfvh$ and $\gfvphi$:
\beq
\gfvh= \left(d+\gamma b\right)\,,\quad
\gfvphi = \left(c+\gamma  a\right)\,.
\label{gdefs}
\eeq
In contrast,
the $gg$ and $\gam\gam$ couplings of the $h$ and $\phi$ come from two
sources: (1) the standard loop contributions computed using
the above 
$f\anti f/VV$ strength factors $\gfvh$ or $\gfvphi$; and (2)
``anomalous'' contributions 
which are expressed in terms of 
the SU(3)$\times$SU(2)$\times$U(1) $\beta$ function coefficients.
The complicated dependence of $\gfvh^2$ and $\gfvphi^2$ on
$\xi$ and $\mphi$ is shown in Fig.~\ref{couplingshmh120} for
$\mh=120\gev$ and $\lphi=5\tev$.
Note that if
$\gfvh^2<1$ is observed, then $\mphi>\mh$, and vice versa,
except for a small region near $\xi=0$. Also note that the radion coupling
$\gfvphi^2$ is generally 
rather small and exhibits zeroes; however, if $\mphi>\mh$ then
at large $|\xi|$ the $ZZ\phi$ couplings can become
sort of SM strength, implying SM type discovery modes could become
relevant (see \cite{Dominici:2002jv}).

A few notes on branching ratios (see \cite{Dominici:2002jv}).
The $h$ branching ratios are quite SM-like (even if partial widths
are different) except that $h\to gg$ can be bigger than normal,
especially when $\gfvh^2$ is suppressed.
For $\mphi<2\mw$, $\phi\to gg$ is very possibly the dominant mode in
the substantial regions near zeroes of $\gfvphi^2$.
However, for
$\mphi>2\mw$ the $\phi$ branching ratios are sort of SM-like
(except at $\xi\simeq 0$) but total and partial widths are rescaled.

We now turn to the LHC, ILC and \gamc\ capabilities. 
We will focus entirely on the case of $\mh=120\gev$. For
the LHC and ILC, we summarize the work of
Ref.~\cite{Battaglia:2003gb}. For the LHC, we rescaled the statistical significances predicted for
the SM Higgs boson at the LHC using $\gfvh^2$ or $\gfvphi^2$
and the modified branching ratios.
We found that
the most important modes for Higgs-radion discovery 
are $gg\to\h\to\gam\gam$ 
and $gg\to \phi\to ZZ^{(*)}\to 4\ell$.
Also useful are $t\anti t \h$ with $\h\to b\anti b$ and $gg\to\h\to ZZ^*\to 4\ell$.
\begin{figure}[h]
\includegraphics[height=1.65in]{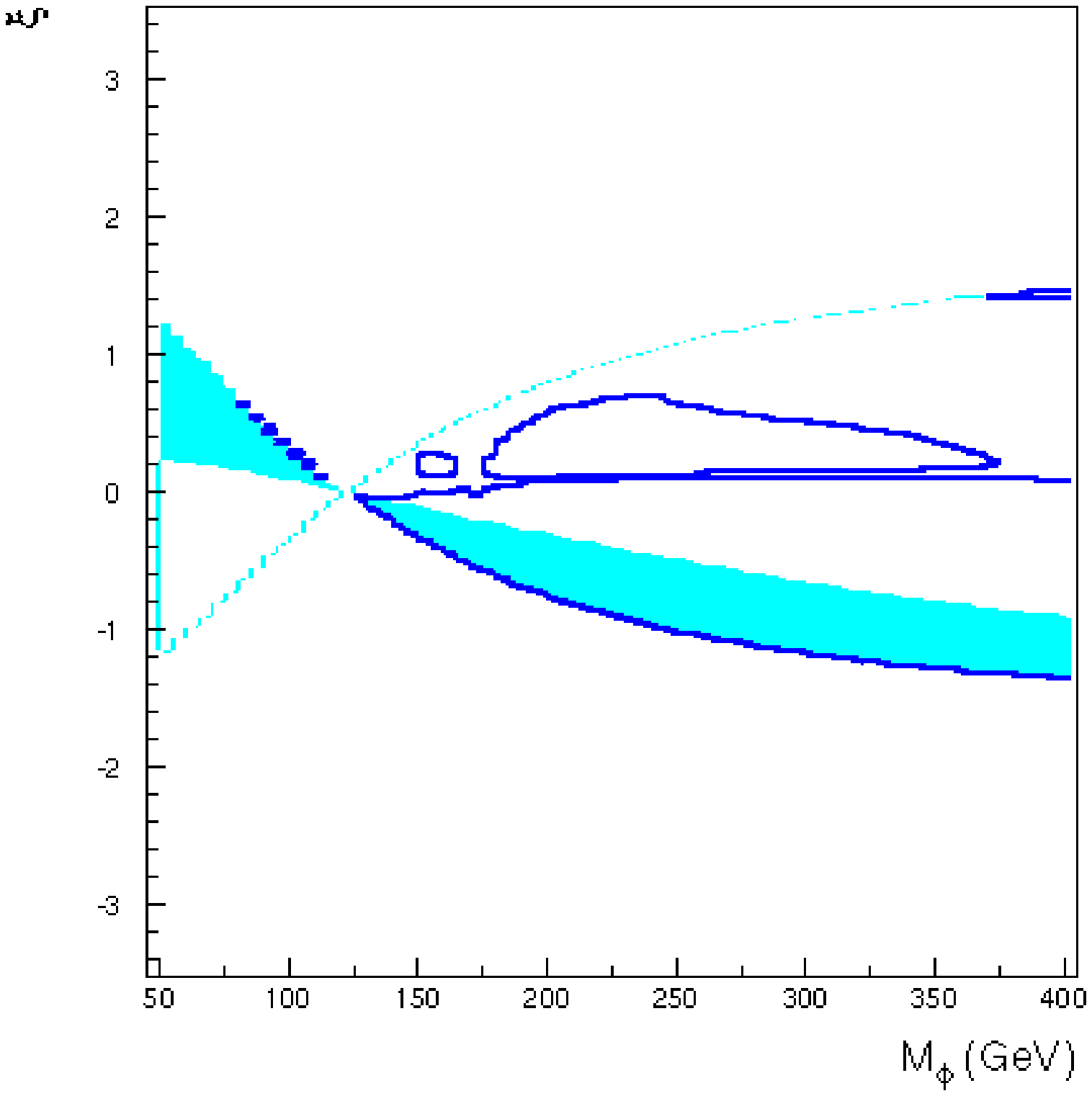}\hspace*{-.2in}
\includegraphics[height=1.65in]{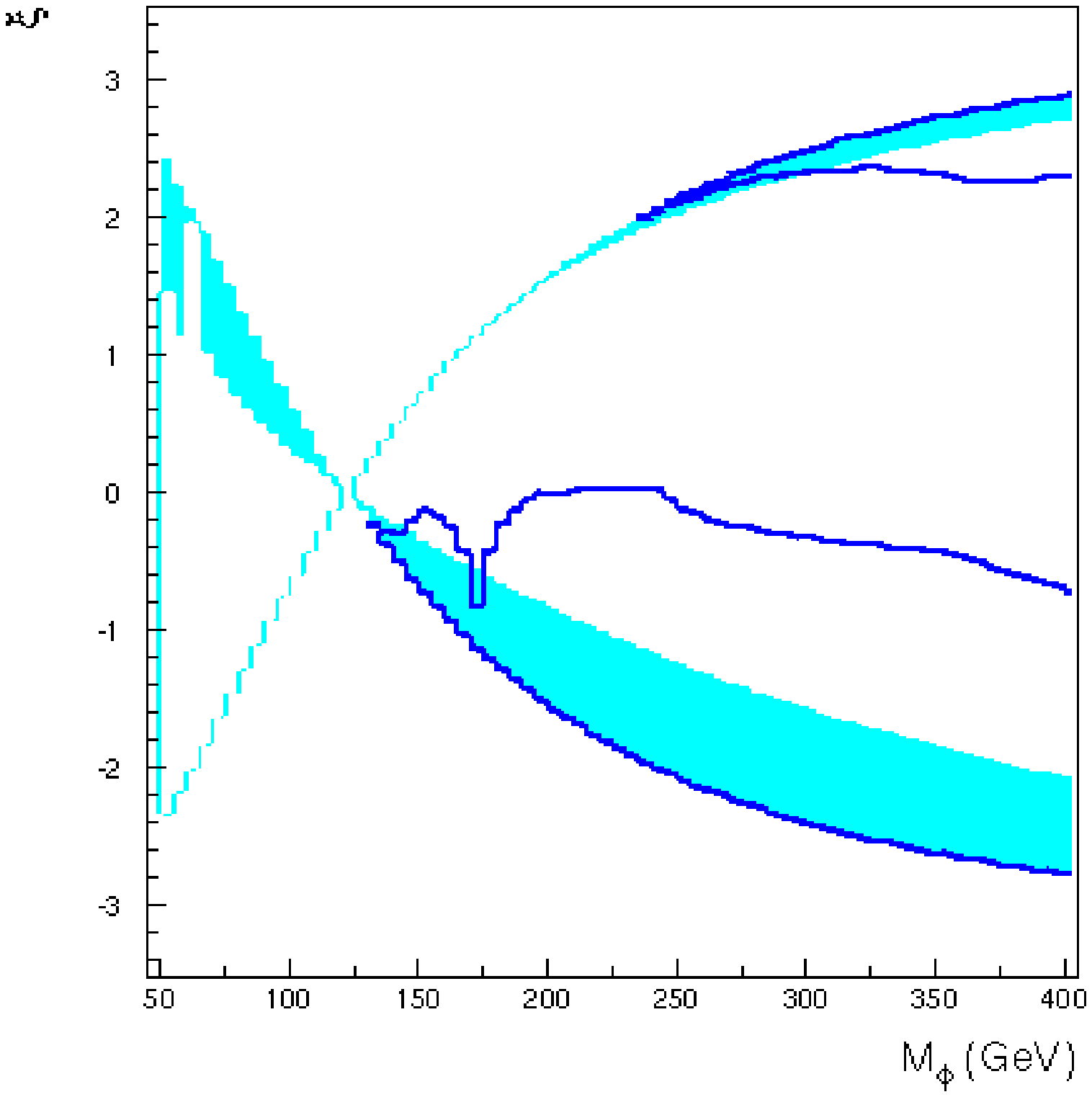}
\hspace*{-.2in}
\includegraphics[height=1.65in]{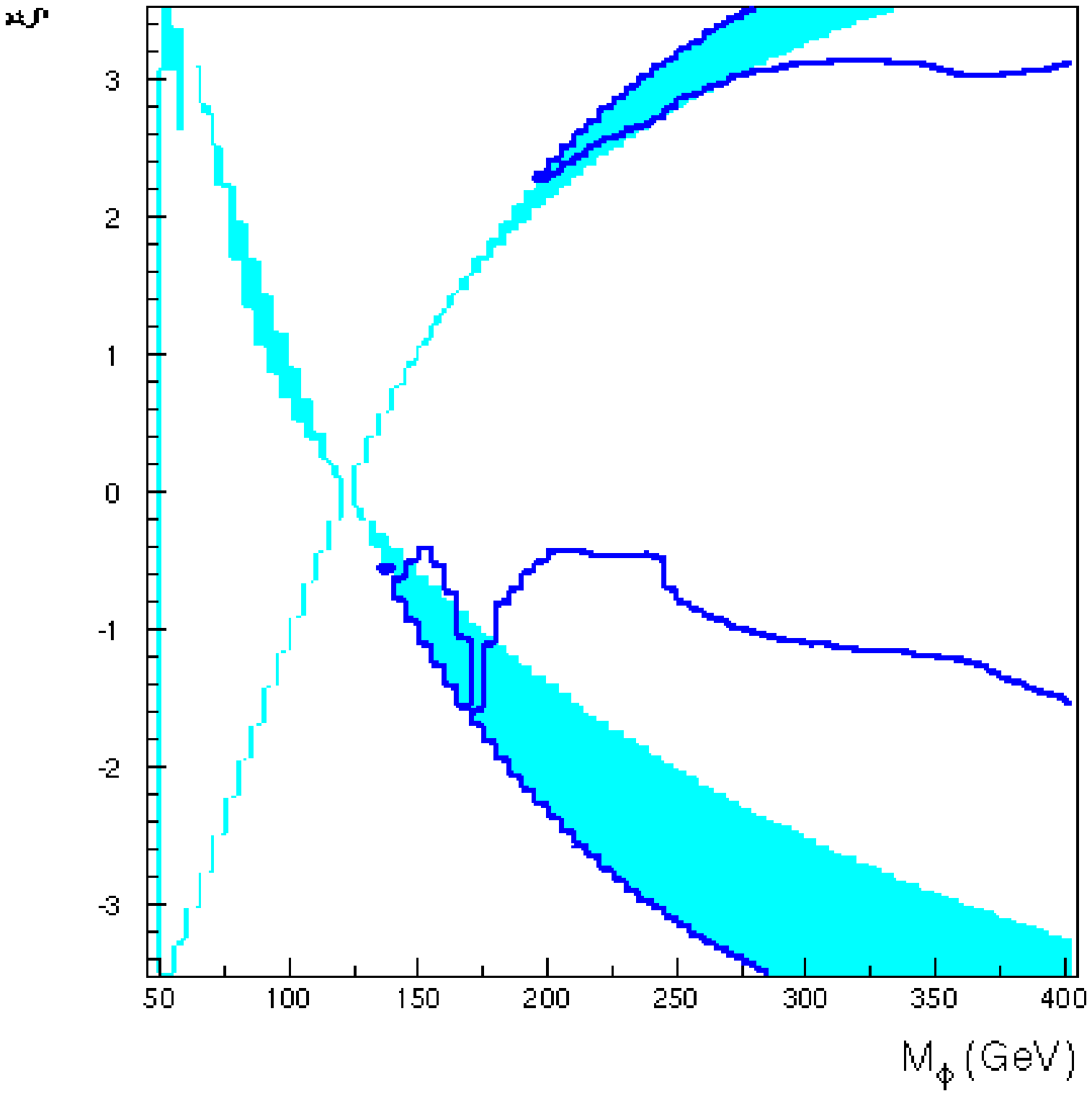}
\vspace*{-.17in}
\caption{ We consider $L=30\fbi$ at  the LHC for $\mh=120\gev$.
The hour-glass outer boundaries define the theoretically consistent
parameter region. Within these, 
the blank (white) regions are where neither the $gg\to\h\to\gam\gam$ mode
nor the $gg\to\h\to 4\ell$ mode yields a $>5\sigma$ signal.  
The regions between dark blue curves are
where $gg\to\phi\to 4\ell$ is $>5\sigma$.
The graphs are for $\lphi=2.5\tev$ (left) $\lphi=5\tev$ (center) and
$\lphi=7.5\tev$ (right). From~\protect\cite{Battaglia:2003gb}.}
\label{fig:compl120}
\vspace*{-.19in}
\end{figure}
Fig.~\ref{fig:compl120} summarizes our results.
It shows that the LHC can find either the $\h$ or $\phi$
unless $\mphi<\mh$ and $\xi>0~\mbox{ and large}$. 
\footnote{However, $|\xi|\lsim 1.5$ is preferred by precision data in
the $\lphi=5\tev$ case.} The region where neither
the $\h$ nor the $\phi$ can be detected 
grows (decreases) as $\mh$ decreases (increases).
It diminishes as $\mh$
increases since the $gg\to \h\to 4\ell$ rate increases at higher $\mh$.
The regions where the $\h$ is not observable
are reduced by considering either 
a larger data set or $qqh$ Higgs production, in association with forward jets. 
Figure~\ref{fig:compl120} also exhibits regions at large $|\xi|$ with $\mphi>\mh$
in which {\it both} the $h$ and $\phi$
mass eigenstates will be detectable.
In these regions, the LHC will observe two scalar bosons
somewhat separated in mass, with the lighter (heavier) having a non-SM-like 
rate for the $gg\to h\to \gamma\gamma$ ($gg\to \phi\to ZZ$) final state.
Additional information will be required to ascertain whether these two Higgs
bosons derive from a multi-doublet or other type of extended
Higgs sector or from the present type of model with Higgs-radion
mixing. For this, we must turn to the ILC and \gamc.

At an $e^+e^-$ ILC, any light scalar, $s$, will be detected in the
$Z^*\to Z s$ mode if $g_{ZZs}^2/g_{ZZ\hsm}^2\gsim 0.01$.  
Since $g_{ZZh}^2/g_{ZZ\hsm}^2=\gfvh^2\geq
0.2$ throughout all of the allowed parameter region, see
Fig.~\ref{couplingshmh120}, observation of the $h$ at the ILC is
guaranteed.  In contrast, Fig.~\ref{couplingshmh120} shows that
$\gfvphi^2\leq 0.01$ for a
significant part of parameter space (smaller $|\xi|$, especially when $\mphi>\mh$).
Unfortunately, as shown in Ref.~\cite{Battaglia:2003gb}, this is also the region where precision measurements of
the $h$ properties at the ILC will deviate by $\lsim 2.5\sigma$
from SM expectations and we could mistakenly conclude
 that the Higgs sector was that of the SM.
%

Can a $\gam\gam$ collider at the ILC or $\gam\gam$ collider
based on a few CLIC modules help? 
To assess, we recall the results for the 
SM Higgs boson obtained in the CLIC study of \cite{Asner:2001vh}.
There, a SM Higgs boson with $\mhsm=115\gev$ was examined.
After cuts, one obtains signal and background rates of $S=3280$
and $B=1660$ 
in the $\gam\gam\to \hsm\to b\anti b$ channel, corresponding to
$S/\sqrt B \sim 80$!

First, consider the $h$. By rescaling to obtain $S_h$ from $S_{\hsm}$,
one finds that the $\gam\gam\to \h\to b\anti b$ rate  
is either changed very little or somewhat enhanced for $\mphi<\mh$
and only modestly suppressed for $\mphi>\mh$ (\eg\ a factor of 2 at
$\mphi=200\gev$). 
Thus, at worst, we would have
$S_h/\sqrt B\sim \half 80 \sim 40$, which
is still a very strong signal.
In fact, we can afford a reduction by a factor
of $16$ before we hit the $5\sigma$ level!  
Thus, {\it the $\gam\gam$
collider will allow $h$ discovery (for $\mh=120$) 
throughout the entire hourglass},
which is something the LHC cannot absolutely do.
In contrast, using the factor of $16$ mentioned above,
the $\phi$ with $\mphi<120\gev$ is very likely to
elude discovery in the $\gam\gam\to\phi\to b\anti b$ mode.
For the $\mphi>\mh$ region, $\gam\gam\to \phi\to WW,ZZ$
would be the best mode, but our current results are not encouraging.

\begin{figure}[h!]
\begin{center}
\includegraphics[height=4.5in,width=2.7in,angle=90]{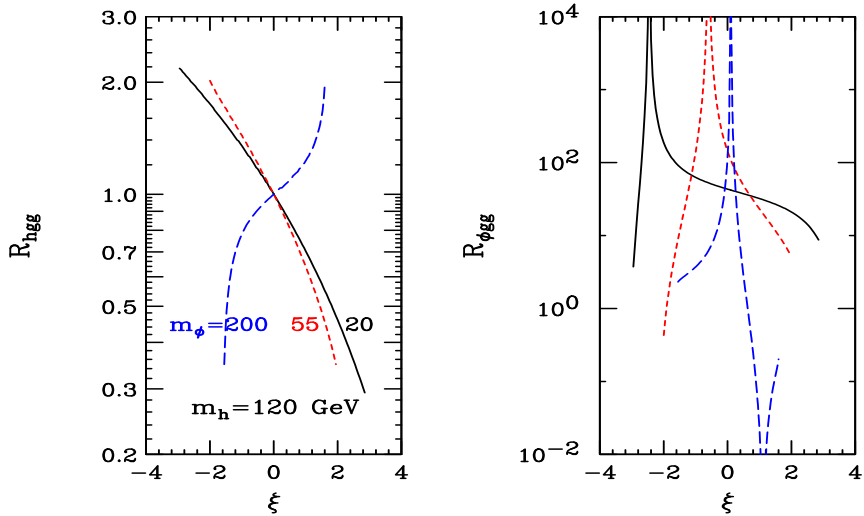}
\end{center}
\vspace*{-.1in}
\caption{ We plot the ratios $R_{\h gg}$ and $R_{\phi gg}$ obtained
after including the anomalous $ggh$ and $gg\phi$, respectively, coupling contributions.
Results are shown
for $m_h=120\gev$  and $\lphi=5\tev$ as functions of $\xi$
for $\mphi=20$, $55$ and $200\gev$.
(The same type of line is used for a given $\mphi$ in 
the right-hand figure as is used in the left-hand figure.) 
From~\protect\cite{Dominici:2002jv}.} 
\label{gggagaanomalouscoup_mh120}
\vspace*{-.2in}
\end{figure}

It is important to emphasize that the \gamc\ can play a very
special role even if we only observe the $h$ there.
Indeed, let us suppose that the $\phi$ is not seen
at any of the three colliders.
The $h$ is very likely to be seen at the
  LHC for $L>100\fbi$ and, as discussed, will be seen
at the \gamc\ and the ILC.
Since $\mh$ will be well-measured, only $\mphi$ and $\xi$ need to
be determined ($\lphi$ having been determined as outlined earlier).
This requires two measurements, with three
or more measurements needed to test the model.
If we could trust LHC and \gamc\  and ILC
absolute rates
(systematics being the question), their different dependencies
on the parameters imply that we
could then determine $\mphi$ and $\xi$ and test the model
even if we don't see the $\phi$.
An interesting way to phrase the LHC and \gamc\  rate measurements
is in terms of the ratio of the rates:
${Rate(gg\to \h \to \gam\gam)\over Rate(\gam\gam\to\h\to b\anti b)}=
{ {\Gamma(gg\to h)\Gamma(h\to \gam\gam)\over \Gamma_h^{tot}}
\over {\Gamma(\gam\gam \to h)\Gamma(h\to b\anti b)\over
  \Gamma_h^{tot}} }=  {\Gamma(\h\to gg)\over \Gamma(\h\to b\anti b)}\,.
$
Using this ratio, we may compute
\beq
R_{hgg}\equiv \left[{\Gamma(\h\to gg)\over \Gamma(\h\to b\anti b)}\right]
 \left[{\Gamma(\hsm\to gg)\over \Gamma(\hsm\to b\anti b)}\right]^{-1}\,,
\eeq
which is the most {\it direct} probe
for the presence of the anomalous $gg h$ coupling~\cite{Asner:2002aa}.
In particular, 
$R_{hgg}=1$ if the only contributions to $\Gamma(\h\to gg)$ come
from quark loops and all quark couplings scale in the same way.
Since the RS model predicts anomalous $gg$ coupling
  contributions
in addition to rescaled standard loop contributions,
substantial deviations from $R_{hgg}=1$  are predicted, as shown in Fig.~\ref{gggagaanomalouscoup_mh120}.

We can estimate the accuracy with which $R_{hgg}$ can be measured
as follows.  
Assuming the maximal reduction of $S_h/S_{\hsm}=1/2$, we find that
$\Gamma(\h\to \gam\gam)\Gamma(\h\to b\anti b)/
\Gamma_{tot}^h$ can be measured
with an accuracy of about $\sqrt{S_h+B}/S_h\sim \sqrt{3200}/1600\sim 0.035$.
The dominant error will then be from the LHC which will typically
measure $\Gamma(\h\to gg)\Gamma(\h\to \gam\gam)/\Gamma_{tot}^h$
with an accuracy of between $0.1$ and $0.2$ (depending on
parameter choices and available $L$).  
{From Fig.~\ref{gggagaanomalouscoup_mh120}, we see that $0.2$
fractional accuracy will reveal deviations 
of $R_{hgg}$ from $1$ for all but the smallest $\xi$ values.}
Given the measured $\mh$,
the direction and magnitude of those deviations will give
a strong constraint on $\mphi$ relative to $\xi$ (although,
for instance, 
you can't tell if $\mphi<\mh$ and $\xi<0$ or $\mphi>\mh$ and $\xi>0$).

Now suppose we also observe the $\phi$. If $|\xi|$ is large,
this is possible at the ILC for any $\mphi$
(see the $\gfvphi^2=0.01$ contour of Fig.~\ref{couplingshmh120}) 
and at the LHC if $\mphi>\mh$ (see
Fig.~\ref{fig:compl120}).
The value of $R_{hgg}$ combined with knowing $\mphi$
will then determine $\xi$ without relying on any absolute rates.
In addition, the  
$\epem\to Z^*\to Z\phi$ rate will have reliable absolute normalization
and it directly determines 
$
{g_{ZZ\phi}^2/ g_{ZZ\hsm}^2}=\gfvphi^2\,.
$
Since $\gfvphi^2$ is wildly varying as a function of the model
parameters (see Fig.~\ref{couplingshmh120}), its measured value
will over constrain and test the model.
If the {LHC} also sees the $\phi$ we get
the model-testing $gg\to\phi\to ZZ$ rate, leading to
a further cross check on the model.

We summarize assuming that $\lphi\lsim 20\tev$. First, $\lphi$
will be measured from the KK $m_{jj}$ spectrum at the LHC. Further,
for such $\lphi$, the
\gamc, like the ILC, can see a light $h$ for all of the $(\xi,\mphi)$ RS parameter space.
Both colliders can see the $h$ where the LHC can't, although the
``bad'' LHC regions are not very big for full $L$.  The ability to
measure $R_{hgg}$ may be the strongest reason for having the \gamc\ as
well as the LHC and ILC, not only in the RS context but also since
most non-SM Higgs theories predict $R_{hgg}\neq 1$ for one reason
or another, unless one is in the decoupling limit.  Further, if the
$\phi$, as well as the $h$, is detected at the ILC, the motivation for
building the \gamc\ becomes even stronger since the measured values of
$\mh$, $\mphi$, $R_{hgg}$ {\it and} $\gfvphi^2$ provide a very
definitive over constrained test of the RS model.   If
$\mphi>\mh$ and $|\xi|$ is large enough for 
detection of $gg\to \phi \to ZZ$ at
the LHC to be possible, the ILC would not be critical (but the \gamc\
would be) since we
could get a definitive determination of $\xi$ using the
measured $\mh$, $\mphi$ and $R_{hgg}$ values and then the $gg\to \phi\to ZZ$ rate
would test the model.  Further model tests would be possible if we
could accurately measure the rate for $h$ production in other LHC
and/or \gamc\ channels --- something that is certainly possible, but
not guaranteed (especially with high accuracy).  Overall, there is a
nice complementarity among the machines --- each brings new abilities
to probe and definitively test the scalar sector of the  RS model.  Very generally, the 
case for a (low-energy) $\gam C$ is compelling if a Higgs
boson is seen at the LHC that has non-SM-like rates and properties.

\vspace*{-.20in}
\section*{Acknowledgment}
\vspace*{-.13in}
This review derives from work in collaboration with
D. Asner, M. Battaglia, S. de Curtis, A. De Roeck, D. Dominici,
J. Gronberg, B. Grzadkowski, M. Velasco, M. Toharia, and J. Wells and
was supported by the U.S. Department of Energy.


\end{document}